\documentclass[a4paper,twocolumn,DIV=14]{scrartcl}
\pdfoutput=1

\usepackage[utf8]{inputenc}
\usepackage[T1]{fontenc}
\usepackage{stix}

\usepackage{scrlayer-scrpage}
\lohead{M Pflüger, RJ Kline, A Fernández Herrero, M Hammerschmidt, V Soltwisch, M Krumrey: Extracting Dimensional Parameters of Gratings Produced with Self-Aligned Multiple Patterning Using GISAXS}
\usepackage{sidecap}
\sidecaptionvpos{figure}{t}

\usepackage{amsmath,amsfonts}
\usepackage{graphicx}
\usepackage[separate-uncertainty=true]{siunitx}
\usepackage{booktabs}
\usepackage{abstract}

\usepackage[format=plain,indention=5pt,labelfont=bf,figurename=Figure,tablename=Table]{caption} % nicer caption

\usepackage[protrusion=true,expansion=true]{microtype} % some improvements for typography

\usepackage[backend=bibtex, style=authoryear, natbib=true,url=false]{biblatex}
\bibliography{report}   % bibliography data in report.bib
\renewcommand{\cite}{ \citep}

\makeatletter% because of @ at preamble code
\renewcommand*{\@fnsymbol}{\@arabic}
\makeatother% because of \makeatletter

\usepackage{hyperref}

\hypersetup{
    pdfauthor={Mika Pflüger, R. Joseph Kline, Analía Fernández Herrero, Martin Hammerschmidt, Victor Soltwisch, Michael Krumrey},     % author
    pdfkeywords={GISAXS, Grazing Incidence Small Angle X-Ray Scattering, line profile, reconstruction, pitchwalk, lithography},
    pdftitle={Extracting Dimensional Parameters of Gratings Produced with Self-Aligned Multiple Patterning Using GISAXS},
    pdfsubject={In this paper, GISAXS is used to reconstruct the line profile of a grating with 32 nm pitch produced by self-aligned quadruple patterning (SAQP). It is found that the reconstructed shape is generally in agreement with previously published SAXS results. However, the reconstructed line height and line width show deviations of 1.0(2) nm and 2.0(7) nm, respectively. Additionally, a series of grating samples with deliberately introduced pitchwalk was measured. Here, it is found that GISAXS yields the pitchwalk in agreement with the SAXS results, with uncertainties ranging from u < 0.5 nm for small pitchwalks (< 2 nm) up to u ≈ 2 nm for larger pitchwalks.},
    colorlinks=true,       % false: boxed links; true: colored links
    linkcolor=black,          % color of internal links
    citecolor=black,          % color of citation anchors
    filecolor=black,          % color of links to local files
    urlcolor=blue,            % color for external urls
    menucolor=black,          % whatever
}

\usepackage{doi}

\newcommand{\ZZ}{\mathbb{Z}}

\renewbibmacro*{doi+eprint+url}{%
    \printfield{doi}%
    \newunit\newblock%
    \iftoggle{bbx:eprint}{%
        \usebibmacro{eprint}%
    }{}%
    \newunit\newblock%
    \iffieldundef{doi}{%
        \usebibmacro{url+urldate}}%
        {}%
    }

\begin{document} 

\title{Extracting Dimensional Parameters of Gratings Produced with Self-Aligned Multiple Patterning Using GISAXS}

\author{Mika Pflüger\thanks{Contact: mika.pflueger@ptb.de}\, \thanks{Physikalisch-Technische Bundesanstalt (PTB), Abbestraße 2-12, 10587 Berlin, Germany} \and R. Joseph Kline\thanks{National Institute of Standards and Technology, Materials Measurement Laboratory, Gaithersburg, Maryland, United States} \and Analía Fernández Herrero\footnotemark[2] \and Martin Hammerschmidt\thanks{JCMwave GmbH, Bolivarallee 22, 14050 Berlin, Germany} \and Victor Soltwisch\footnotemark[2] \and Michael Krumrey\footnotemark[2]}

\twocolumn[
\maketitle

\begin{onecolabstract}
% The abstract should consist of a single paragraph containing no more than 200 words. It should be a summary of the paper and not an introduction. Because the abstract may be used in abstracting and indexing databases, it should be self-contained (that is, no numerical references) and substantive in nature, presenting concisely the objectives, methodology used, results obtained, and their significance.
% DO NOT BELIEVE WHAT IS WRITTEN ABOVE! IT IS NOT TRUE ANYMORE! ACTUALLY, JM3 WANTS INSTEAD (they just forgot to tell us):
% Each subsection should contain 1-2 sentences, answering the following questions:
% Background: What issues led to this work? What is the environment that makes this work interesting or important?
% Aim: What did you plan to achieve in this work? What gap is being filled?
% Approach: How did you set about achieving your aims (e.g., experimental method, simulation approach, theoretical approach, combinations of these, etc.)? What did you actually do?
% Results: What were the main results of the study (including numbers, if appropriate)?
% Conclusions: What were your main conclusions? Why are the results important? Where will they lead?
% 200 word limit of course still applies, what did you think?

\textbf{Background:} To ensure consistent and high-quality semiconductor production at future logic nodes, additional metrology tools are needed.
For this purpose, grazing-incidence small-angle X-ray scattering (GISAXS) is being considered because measurements are fast with a proven capability to reconstruct average grating line profiles with high accuracy.

\textbf{Aim:} GISAXS measurements of grating line shapes should be extended to samples with pitches smaller than \SI{50}{nm} and their defects.
The method's performance should be evaluated.

\textbf{Approach:} A series of gratings with \SI{32}{nm} pitch and deliberately introduced pitchwalk is measured using GISAXS\@.
The grating line profiles with associated uncertainties are reconstructed using a Maxwell solver and Markov-Chain Monte Carlo (MCMC) sampling combined with a simulation library approach.

\textbf{Results:} The line shape and the pitchwalk are generally in agreement with previously published transmission small-angle X-ray scattering (SAXS) results;
however the line height and line width show deviations of \SI{1.0(2)}{nm} and \SI{2.0(7)}{nm}, respectively.
The complex data evaluation leads to relatively high pitchwalk uncertainties between \SI{0.5}{nm} and \SI{2}{nm}.

\textbf{Conclusions:} GISAXS shows great potential as a metrology tool for small-pitch line gratings with complex line profiles.
Faster simulation methods would enable more accurate results.
\end{onecolabstract}
\bigskip
]

\saythanks

% A list of up to six keywords should immediately follow, with the keywords separated by commas and ending with a period. The body of the manuscript should be double-spaced and fully justified.
%\keywords{GISAXS, X-ray Scattering, pitchwalk, metrology, grating, SAQP}

\section{Introduction}
\label{sec:intro}

To manufacture semiconductor structures with dimensions smaller than the Abbe limit\cite{abbe_1873_BeitraegeZur}, multi-patterning methods have been developed\cite{hazelton_2009_DoublepatterningRequirements}, including self-aligned multiple patterning\cite{jung_2006_PatterningSpacer}.
Self-aligned double patterning (SADP) as well as self-aligned quadruple patterning (SAQP) have already been introduced into high volume manufacturing for the "14nm" node \cite{natarajan_2014_14nmLogic} and for the "10nm" node\cite{yeoh_2018_InterconnectStack}, respectively.
These manufacturing schemes entail complex geometries, and errors in early processing steps can be propagated in later steps and lead to additional defects.
Therefore, metrology tools suited for the measurement of these complex geometries need to be developed\cite{orji_2018_MetrologyNext}.

Because no method can address all metrology needs alone, imaging methods which excel in the study of small areas and individual features have to be developed as well as statistical methods with a large field of view which deliver average information over a large structured area.
For imaging applications, scanning electron microscopy (SEM) as well as atomic force microscopy (AFM) are widely employed in the industry and there is active research to further develop them for future metrology needs.
Additionally, other probes for microscopy are investigated, including helium ion microscopy and proton microscopy\cite{bunday_2018_75nmLogic}.
For statistical, non-imaging applications, the industrially established method is scatterometry including optical critical dimension metrology (OCD);
Current research is focused on extensions such as Mueller-matrix spectroscopic ellipsometry (MMSE)\cite{novikova_2006_ApplicationMueller,dixit_2015_MetrologyBlock,diebold_2018_PerspectiveOptical} and virtual reference OCD\cite{vaid_2015_ImprovedScatterometry,bunday_2018_75nmLogic}.
However, new methods with high statistical power are also investigated and the most important is small-angle X-ray scattering (SAXS)\cite{jones_2003_SmallAngle,bunday_2018_75nmLogic}.

SAXS has distinct advantages, but there are also several challenges which have so far hindered industrial applications.
Like optical scatterometry, X-ray scattering probes the average structure with \si{nm} precision over a relatively large (\si{\micro\meter\squared}) area\cite{sunday_2015_DeterminingShape}.
Also like in optical scatterometry, modeling has to be used to determine relevant structural parameters from a SAXS measurement, but there are two key advantages compared to optical measurements.
Firstly, because the employed wavelength is much smaller than the feature sizes multiple diffraction orders can be measured;
the additional information aids the reconstruction, such that complex models with more than 12 parameters can be reconstructed, parameter correlations can be reduced, and unique solutions can be determined more easily\cite{sunday_2015_DeterminingShape,bunday_2018_75nmLogic}.
Secondly, the refractive index of X-rays far from the elements' absorption edges only depends on the elemental composition (mostly, the electron density) and the refractive indices are well known\cite{henke_1993_XrayInteractions}.
This enables reference-free modeling of the SAXS measurements and consequently the evaluation of uncertainties and traceability of the measurement results to the international system of units (SI) without the need for reference measurements\cite{sunday_2016_EvaluationEffect}.
The main limitation of transmission SAXS are the small signal intensities and consequently long measurement times.
SAXS signal intensities are small because the brightness of X-ray sources is limited, X-rays are attenuated during transmission through the wafer, and the scattering cross section of X-rays with matter is small.

Signal intensities of X-ray scattering measurements can be enhanced significantly by measuring in reflection geometry at small incidence angles (grazing-incidence small-angle X-ray scattering, GISAXS), which avoids transmission through the wafer and additionally enhances scattering because of total external reflection at the critical angle\cite{levine_1989_GrazingincidenceSmallangle}.
Therefore, measurements in reflection geometry approach acceptable measurement speed\cite{bunday_2018_75nmLogic}.
As a consequence of the small incidence angles, the X-rays do not penetrate deeply into the sample, so that GISAXS is a surface-sensitive technique.

Some additional challenges arise from the GISAXS geometry.
Because the small incidence angle leads to an elongation of the X-ray beam on the sample, very large (\si{\milli\meter\squared}) areas are probed at the same time.
While this is acceptable or even desirable in applications with large homogeneous sample structures such as memory manufacturing\cite{hagihara_2017_CapabilityMeasuring,hagihara_2019_ThroughputImprovement}, other applications rely on the measurement of small metrology targets (approximately \SI{50x50}{\micro\meter}).
To achieve smaller beam footprints, the beam spot size and in particular the beam height need to be reduced.
While small beam heights down to \SI{300}{\nano\meter} leading to a footprint on the sample of about \SI{30}{\micro\meter} have been demonstrated, this presents large technical challenges in focusing the beam and aligning the beam on the sample\cite{roth_2007_SituObservation}.
Another approach for the measurement of small metrology targets is to manufacture the targets rotated in-plane with respect to the surrounding logic structures, so that the target scatters to different exit angles than the surrounding logic\cite{pfluger_2017_GrazingincidenceSmallangle,pfluger_2017_SelectiveMeasurement}.

Compared to transmission SAXS measurements, the data evaluation in GISAXS experiments also presents additional challenges.
Multiple scattering processes are common in GISAXS, so that the Born approximation, which is common in SAXS modeling and simplifies the simulation of SAXS measurements considerably, is not applicable in GISAXS\cite{sinha_1988_XrayNeutron}.

Despite these challenges, GISAXS has already been shown to be a suitable method for determining line grating pitches\cite{yan_2007_IntersectionGrating,wernecke_2014_TraceableGISAXS} and line profiles\cite{hofmann_2009_GrazingIncident,soltwisch_2017_ReconstructingDetailed,yamanaka_2016_MeasurementCapabilities} as well as line-edge roughness\cite{suh_2016_CharacterizationShape,fernandezherrero_2019_ApplicabilityDebyeWaller}.
It has also been used to reconstruct the average profile of contact holes\cite{hagihara_2017_CapabilityMeasuring}, to unravel complex hierarchical nanostructures\cite{khaira_2017_DerivationMultiple} and to quantify deviations in nanostructure orientation\cite{pfluger_2019_DistortionAnalysis}.
However, GISAXS reconstructions have been limited to structures with relatively large ($>\SI{50}{nm}$) pitches.
Specifically, measurements of grating structures produced by modern multi-patterning methods, which lead to more complex line profiles and layer stacks, have not been reported to date.

One defect commonly introduced in multi-patterning methods is pitchwalk.
When pitchwalk is present due to alignment errors, the distance between two lines alternates between a higher and a lower value, such that the average pitch stays the same.
Unfortunately, even small changes in the distances between lines can lead to large deviations after further processing steps.
For example, if the lines manufactured by multi-patterning are used as an etch mask for the production of fin structures, small variations in distances can lead to large variations in etch depth\cite{chao_2016_AdvancedInline}.
It was shown that the average pitchwalk of larger areas can be measured using optical scatterometry\cite{diebold_2018_PerspectiveOptical,kagalwala_2016_MeasuringSelfaligned} as well as CD-SAXS\cite{sunday_2015_DeterminingShape}.
Due to the aforementioned advantages of GISAXS, it is desirable to investigate the performance of pitchwalk measurements using GISAXS.

\begin{figure*}[t]
\includegraphics{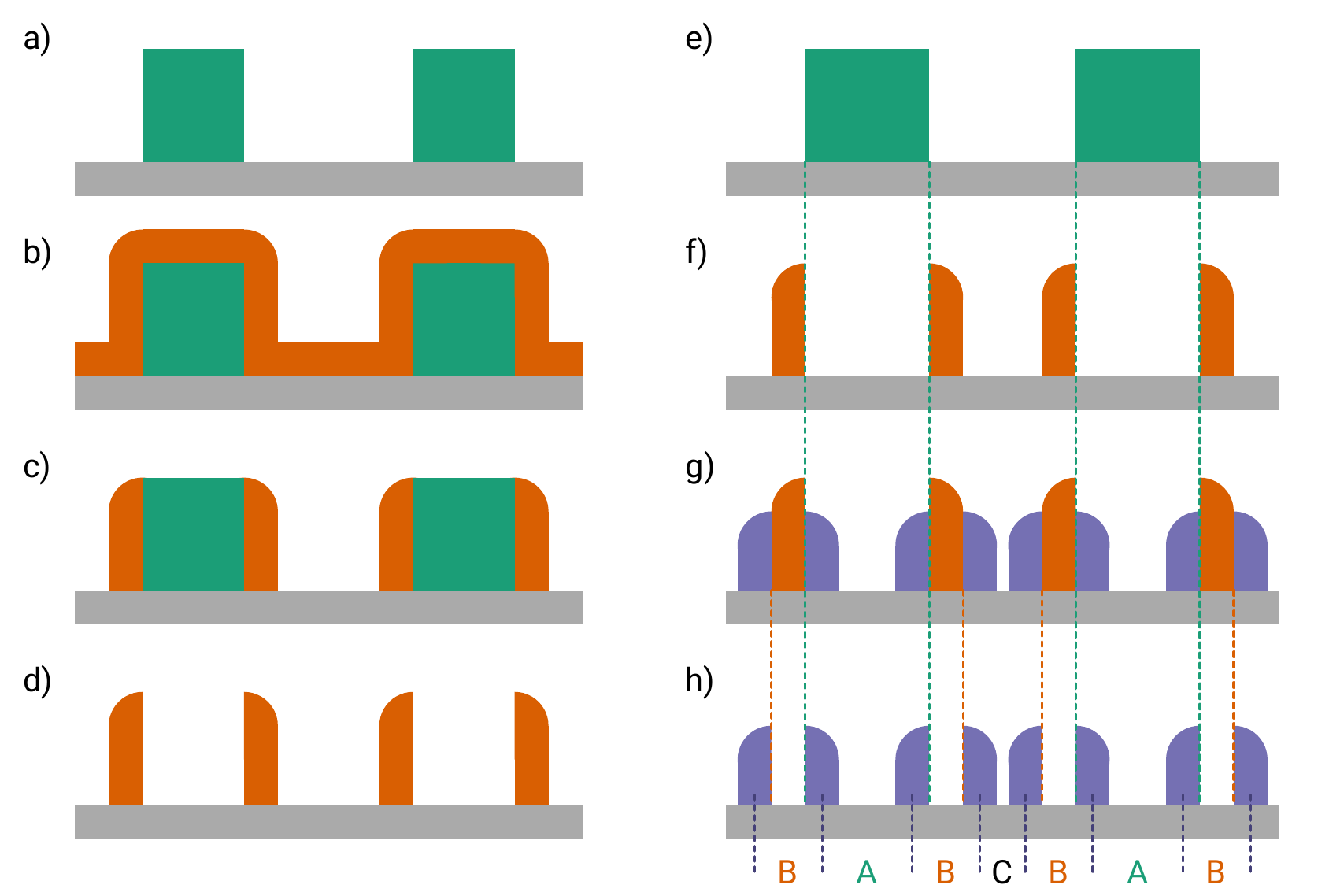}
\caption[Pitchwalk in self-aligned multiple patterning]{
Self-aligned multiple patterning and pitchwalk.
a-d) Self-aligned double patterning principle.
The processing steps are: a) conventional lithography, b) atomic layer deposition, c) anisotropic etching of the deposited layer, leaving only the deposited side walls, and d) chemically selective etching to remove the original line.
e-h) Pitchwalk in self-aligned quadruple patterning.
The panels show processing steps in SAQP leading to a non-zero pitchwalk: e) conventional lithography with a line width that is too large, f) first pitch doubling, g) second pitch doubling, and h) final feature.
Dashed lines show how the distances between lines in the final feature are determined in the process.
The distances $B$ change with the first ALD layer thickness, $A$ changes with the original line width and $C$ changes with the original trench width.
The pitchwalk that was varied in this study is $\delta p = A-C$.}
\label{fig:saqp}
\end{figure*}

In this paper, we will investigate the use of GISAXS for the reconstruction of grating line profiles and for the measurement of pitchwalk introduced by multi-patterning methods.
Specifically, we will reconstruct the line profile of a grating that has a pitch of \SI{32}{nm} and that was manufactured by means of self-aligned quadruple patterning and compare the reconstructed line profile to SAXS measurements qualitatively and quantitatively.
Then, we will report measurements of a series of gratings for which pitchwalk was deliberately introduced.
Finally, we will quantitatively compare the results obtained via GISAXS with results from SAXS measurements.

\section{Methods}

\subsection{Sample Preparation}

\begin{SCfigure*}[1.0][t]
\includegraphics[width=1.3\columnwidth]{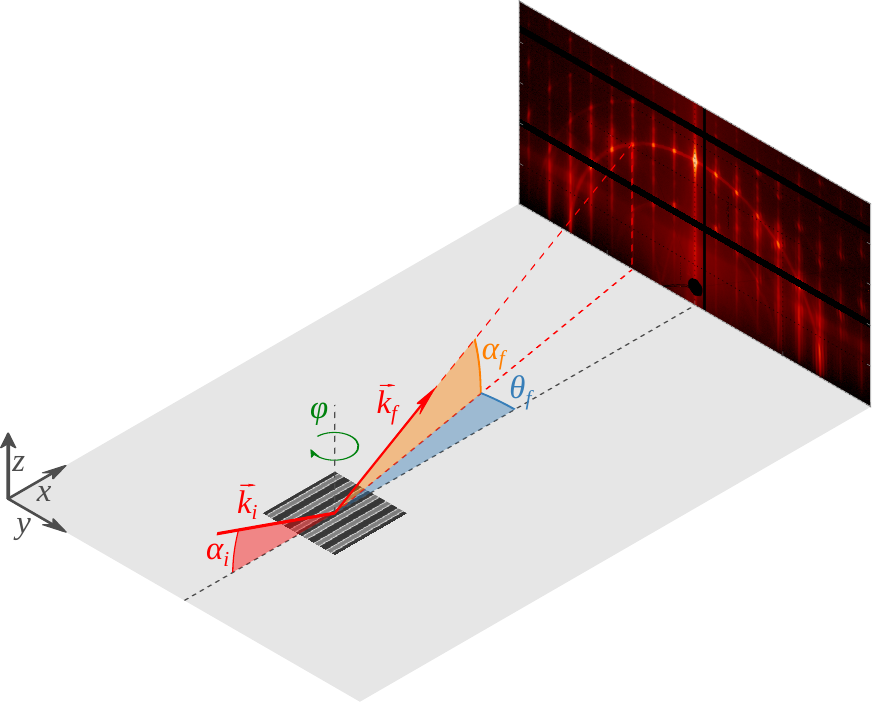}
\caption[Geometry of GISAXS experiments]{
Schematic of a GISAXS experiment.
The incident beam with wave vector $\vec k_i$ impinges under the grazing angle $\alpha_i$ onto the sample surface.
The scattered beam with wave vector $\vec k_f$ and exit angles $\theta_f$ and $\alpha_f$ is collected by an area detector.
The sample can be rotated around the sample normal by the angle $\varphi$.}
\label{fig:gisaxs_geometry}
\end{SCfigure*}

The sample consists of a silicon wafer with measurement targets arranged on it in a regular grid.
Each measurement target covers an area of \SI{1x9}{mm} and other structures are surrounding the measurement targets.
All measurement targets were produced in the same process, which consisted of coating and lithographic exposure followed by etching to produce a line grating with a \SI{128}{nm} pitch and subsequent pitch quartering for a final grating pitch of \SI{32}{nm} using self-aligned quadruple patterning (SAQP) \cite{vanveenhuizen_2012_DemonstrationElectrically,chawla_2014_PatterningChallenges}.
In self-aligned multiple patterning, existing lines are coated uniformly using atomic layer deposition, followed by anisotropic etching, which selectively removes the material deposited on top of the lines and in the trench, leaving the material on the sidewalls (see fig.~\ref{fig:saqp} a-c).
Then, chemically selective etching removes the original lines, leaving the material on the former sidewalls as a grating with a half pitch compared to the original grating (fig.~\ref{fig:saqp} d).
For SAQP, this process is performed twice to reach quarter pitch (fig.~\ref{fig:saqp} e-g).

When the fill ratio of the lithographic exposure, i.e.\ the ratio of line width to pitch, is not correct, self-aligned multiple patterning results in nonuniform line distances;
this defect is referred to as pitchwalk $\delta p = A - C$ (see fig.~\ref{fig:saqp}~h).
The wafer sample was produced with a different lithographic focus and exposure along one axis, resulting in six rows of targets with differing fill ratio and therefore different $\delta p$.
We label the rows PQ~1 – PQ~6 from left to right, where the $\delta p$ grows from negative values at PQ~1 to positive values at PQ~6, crossing nominally zero pitchwalk at PQ~4.
Along the other axis, the production conditions were kept the same, resulting in columns of identical targets.

The grating lines are made of silicon oxide.
Due to the multi-step production process, the grating lines rest on top of a layer structure consisting of \SI{30}{nm} of silicon nitride on top of \SI{25}{nm} of titanium nitride on top of \SI{100}{nm} of silicon oxide, on the silicon wafer.
Further details of the sample production have been published previously \cite{sunday_2015_DeterminingShape,villarrubia_2015_ScanningElectron}.

SAXS\cite{sunday_2015_DeterminingShape,sunday_2016_EvaluationEffect} and electron microscopy\cite{villarrubia_2015_ScanningElectron} measurements of the samples described have already been reported in the literature.
We use the line shape reconstructed from SAXS measurements by Sunday et al.\cite{sunday_2015_DeterminingShape}\ as a comparison for our GISAXS measurements.

\subsection{Grating Diffraction}

A schematic of a GISAXS experiment \cite{levine_1989_GrazingincidenceSmallangle} is shown in fig.~\ref{fig:gisaxs_geometry}.
A flat sample surface is illuminated under a grazing incidence angle $\alpha_i$ using monochromatic X-rays with wave vector $\vec k_i$.
The X-ray beam is scattered elastically according to the geometric features on the sample surface.
The intensity distribution of the scattered X-rays is collected using an area detector, and from the exit angles $\theta_f$ and $\alpha_f$, the wave vector of the scattered beam $\vec k_f$ is calculated.
Considering only elastic scattering, $|\vec k_i| = |\vec k_f| = k = \frac{2\pi}{\lambda}$, with the wave length of the incident light $\lambda$.

For the coordinate system chosen, the sample plane is the $x$-$y$-plane, the $z$-axis is the sample normal and the projection of the incident beam onto the sample plane falls onto the $x$-axis (see fig.~\ref{fig:gisaxs_geometry}).
In this coordinate system, we can express the scattering momentum transfer $\vec q = \vec k_f - \vec k_i$ as
\begin{align}
q_x &= k\, (\cos \theta_f \cos \alpha_f - \cos \alpha_i) \\
q_y &= k\, \sin \theta_f \cos \alpha_f \\
q_z &= k\, (\sin \alpha_i + \sin \alpha_f) \quad .
\end{align}

\begin{figure*}[t]
\includegraphics[width=\textwidth]{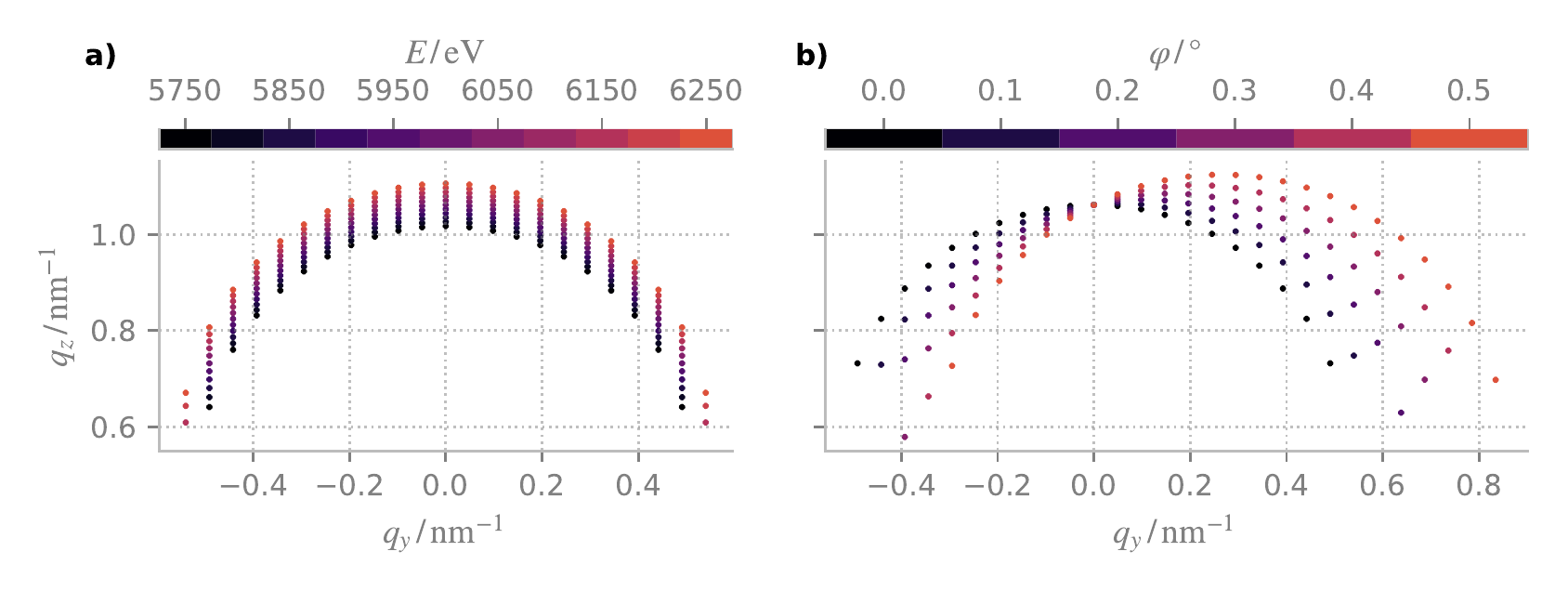}
\caption[GISAXS measurement points in reciprocal space]{
Reciprocal space map of GISAXS measurement points.
a) By varying the photon energy $E$, the diffraction orders are translated in $q_z$.
b) Variations of the azimuthal angle $\varphi$ change the $q_z$ of the higher diffraction orders much more than that of the lower diffraction orders, yielding complementary information.
}
\label{fig:measurement_points}
\end{figure*}

The diffraction of line gratings can be understood qualitatively using reciprocal space construction \cite{mikulik_2001_CoplanarNoncoplanar,yan_2007_IntersectionGrating}.
The three-dimensional Fourier transform of the periodically spaced grating lines consists of grating truncation rods, and their intersections with the Ewald sphere of elastic scattering comprise the diffraction orders.
In the coordinate system adopted in this paper, the position of the grating diffraction orders in reciprocal space is\cite{pfluger_2017_GrazingincidenceSmallangle}:
\begin{align}
    q_x =& \sin \varphi \, n \, 2 \pi / p \\
    q_y =& \cos \varphi \, n \, 2 \pi / p \\
    q_z =& \frac{2 \pi}{\lambda} \bigg(\sin \alpha_i \\
         &+ \sqrt{\sin^2 \alpha_i - (n \lambda / p)^2 - 2 \sin \varphi \cos \alpha_i \, n \lambda / p} \bigg) \, , \nonumber
\end{align}
with the grating pitch $p$, the grating diffraction order $n$ and the sample rotation $\varphi$, with $\varphi=0$ defined in such a way that the projection of the incoming beam onto the sample plane is parallel to the grating lines (conical mounting).
If the grating pitch $p$ equals the unit cell width of the grating, i.e.\ if all lines are identical, the grating diffraction order is an integer, $n \in \ZZ$.
However, in the samples we investigate, the unit cell is a multiple of the pitch, either due to the line shapes being alternated (which doubles the unit cell size) or due to the pitchwalk (which also doubles the unit cell size).
In this case, additional grating diffraction orders arise between the original diffraction orders, and the grating diffraction order assumes half or quarter values, $2n \in \ZZ$ or $4n \in \ZZ$, respectively.

\subsection{GISAXS Experiments}

\begin{figure*}[t]
\includegraphics[width=\textwidth]{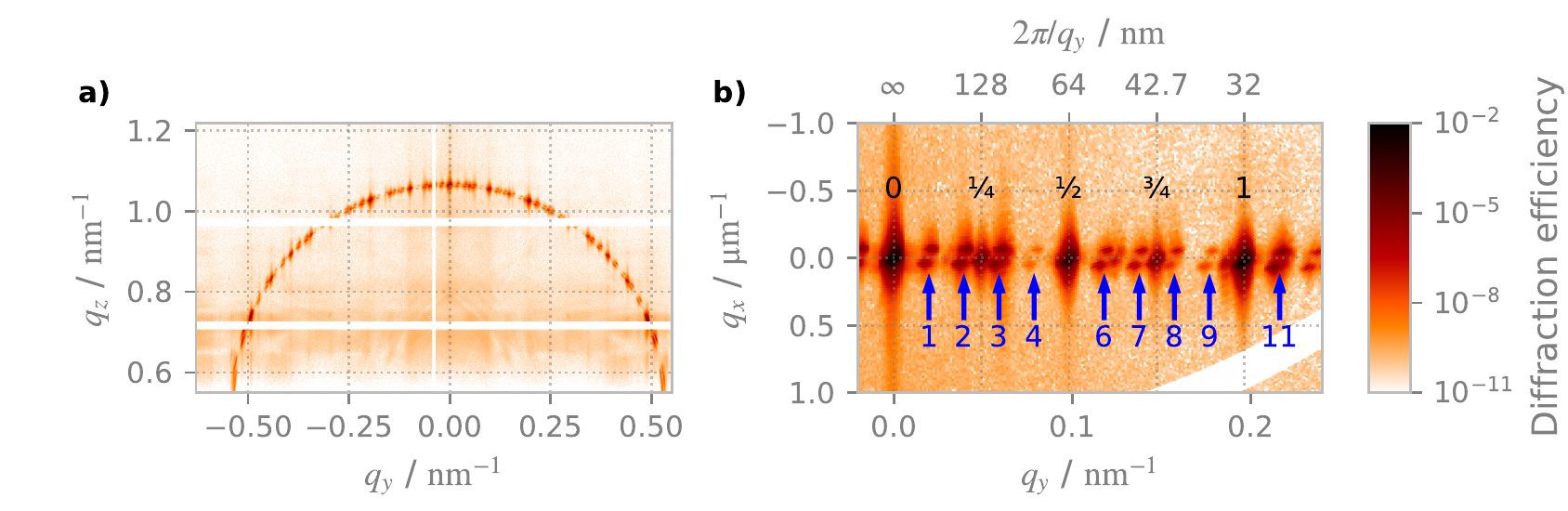}
\caption[Example GISAXS pattern]{
GISAXS pattern example.
a) Scattering pattern taken from the target PQ~1 at $\varphi = \SI{0}{\degree}$ and $E=\SI{6000}{eV}$.
Due to the high quality of the grating, scattering is confined almost exclusively to a semi-circle.
b) Detailed view of the first few diffraction orders, converted to a $q_y$-$q_x$ map.
The diffraction orders of the target grating are marked with black numbers;
the $n=0, \tfrac{1}{4}, \tfrac{1}{2}, \tfrac{3}{4}, 1$ diffraction orders are visible.
The parasitic diffraction orders stemming from the surroundings are additionally marked with blue arrows and numbers.
Each parasitic diffraction order is a double peak (from the surroundings before and behind the target), and due to the \SI{320}{nm} pitch of the surroundings, the $n_{\mathrm{sur}}=1$ diffraction order is at the position where a $n=0.1$ diffraction order of the target grating would be.
GISAXS pattern example.
}
\label{fig:gisaxs_example}
\end{figure*}

GISAXS measurements were performed using the four-crystal monochromator beamline in the laboratory of the Physikalisch-Technische Bundesanstalt \cite{beckhoff_2009_QuartercenturyMetrology} at the BESSY II electron storage ring in Berlin.
The experimental setup consisted of the beamline, a sample chamber and a detector sledge.
The beamline included a monochromator that allowed the photon energy to be adjusted in the range between \SI{1.7}{keV} and \SI{10}{keV} \cite{krumrey_2001_HighaccuracyDetector}, several slits, and two pinhole stages for beam shaping.
The sample chamber \cite{fuchs_1995_HighPrecision} allowed the sample to be positioned in all three directions with a precision of \SI{3}{\micro\meter} and to be rotated around all three axes with a precision of \SI{0.001}{\degree}.
In addition, the Helmholtz-Zentrum Berlin's SAXS detector sledge \cite{gleber_2010_TraceableSize} allows the attached in-vacuum Pilatus 1M\footnote{Certain commercial equipment, instruments, or materials are identified in this paper in order to specify the experimental procedure adequately. Such identification is not intended to imply recommendation or endorsement by any of the authors or the National Institute of Standards and Technology, nor is it intended to imply that the materials or equipment identified are necessarily the best available for the purpose.} area detector \cite{wernecke_2014_CharacterizationInvacuum} to be moved for sample-to-detector distances between \SI{2}{m} and \SI{4}{m} and for exit angles up to approximately \SI{2}{\degree}.
The whole beam path including the sample site is evacuated and a high vacuum is maintained.
For the measurements presented here, the beam spot size was reduced to about \SI{150x150}{\micro\meter} full width at half maximum using a beam-defining \SI{100}{\micro\meter} Pt pinhole (Plano GmbH, Germany) and an adjustable slit system with low-scatter blades (XENOCS, France) as a scatter guard.
At the selected incident angle of approximately $\alpha_i = \SI{1}{\degree}$ for the GISAXS experiments, this results in a size of the projected footprint of about \SI{9}{mm} full width at half maximum.
The GISAXS measurements were taken from a column of targets next to the column measured by SAXS\@.
Because the production conditions were identical in the columns, the measurements are fully comparable.

\begin{figure*}[t]
\includegraphics{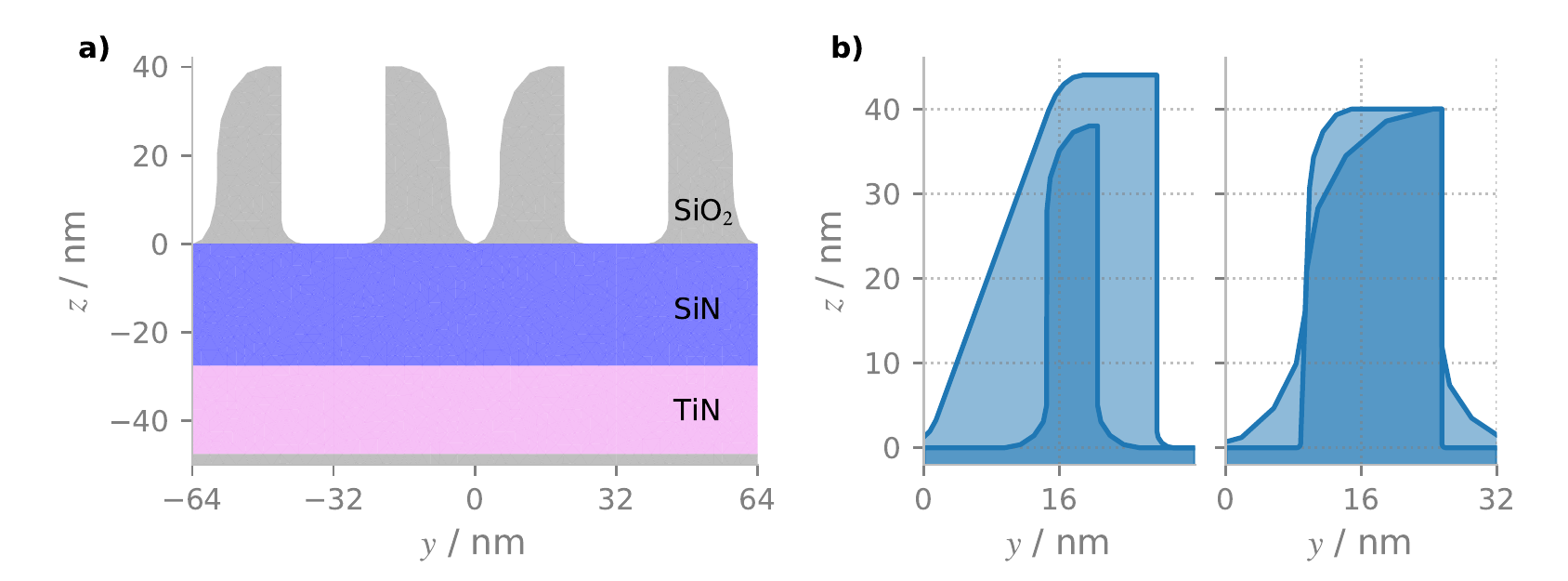}
\caption[Computational model]{
Computational model.
a) Overview of the model and element composition.
Two unit cells are shown.
Thin lines within the structures outline the finite elements used to calculate the electric fields.
For clarity, the thick silicon oxide layer and substrate at the bottom are not shown.
b) Variability of the model.
The boundaries of line width, line height, and side wall angle are shown to the left.
To the right, the boundaries of the corner rounding are shown.}
\label{fig:line_shape_model}
\end{figure*}

GISAXS measurements were taken over a range of X-ray photon energies $E$ and sample rotations $\varphi$.
$\varphi=\SI{0}{\degree}$ was aligned by tuning the sample rotation until the recorded scattering was symmetric along the specular axis, yielding $|\varphi| < \SI{0.005}{\degree}$.
The incident angle was set to approximately $\alpha_i=\SI{1}{\degree}$ and the exact value for $\alpha_i$ was determined from the collected GISAXS patterns and the calibrated sample-to-detector distance.
For all measurement targets, measurements were taken at $\varphi=\SI{0}{\degree}$ for a range of photon energies from \SI{5750}{eV} to \SI{6250}{eV} in steps of \SI{50}{eV}, using $t=\SI{300}{s}$ as the exposure time.
Additionally, at $E=\SI{5900}{eV}$, $E=\SI{6000}{eV}$, and $E=\SI{6100}{eV}$, measurements were taken for a range of sample rotations from $\varphi = \SI{0.1}{\degree}$ to $\varphi = \SI{0.5}{\degree}$ in steps of \SI{0.1}{\degree} using $t=\SI{180}{s}$ (see figure \ref{fig:measurement_points}).
In total, measurements at 26 different $(E, \varphi)$ sets were taken for each target.
Using the signal of a calibrated monitor diode, the scattering images obtained were normalized to the incident flux and the exposure time.
Due to the detector's counting limit of approx.\ \num{1000000} counts per pixel, the dynamic range of the images was enhanced by combining each image with an image with $t=\SI{1}{\second}$.
For this combination, an intensity threshold corresponding to about $\num{1000000}$ counts per pixel in the long exposure time image was used, and all pixels above this threshold were taken from the corresponding $t=\SI{1}{\second}$ image.

From the scattering patterns (see fig.~\ref{fig:gisaxs_example} a) for an example), the intensity of the diffraction orders was extracted by integrating over each diffraction order.
The background noise (mainly from diffuse scattering) was estimated and subtracted from the diffraction orders.
Due to the low incidence angle $\alpha_i$, the projection of the incident beam onto the sample was longer than the measurement target.
Therefore, an additional signal due to scattering of the surroundings of the measurement target is also visible in the scattering patterns.
In the surroundings is a structure with a period of about \SI{320}{nm}, which means the 10th diffraction order of the surrounding structure $n_{\mathrm{sur}}=10$ coincides with the $n=1$ diffraction order of our target with a period of \SI{32}{nm}, see fig.~\ref{fig:gisaxs_example} b).
Fortunately, the scattering of the surrounding structure is strong only for small $n_{\mathrm{sur}}$, meaning that their contribution to the total scattered intensity does not bury the target signal.
To account for these parasitic signals, the intensities of the diffraction orders of the surrounding structure that do not coincide with a diffraction order of our target are extracted for $n_{\mathrm{sur}} \ge 4$.
The effect of the parasitic signals on the diffraction orders is estimated as the mean intensity of the parasitic diffraction orders, which is \num{1.8e-7}.
This is then subtracted from the diffraction intensity of our target where the diffraction orders coincide.
Experimental uncertainties of the coinciding diffraction orders are estimated conservatively as the maximum intensity of the parasitic diffraction orders, which is \num{2.2e-6}.
Experimental uncertainties of the quartered diffraction orders that do not coincide with parasitic diffraction orders ($n = 0.25, 0.75, 1.25$) are estimated as the background noise, which is $< \num{5e-8}$ for all orders.

\subsection{Simulation of the Diffraction Intensity}

\begin{figure}[t!]
\includegraphics[width=\columnwidth]{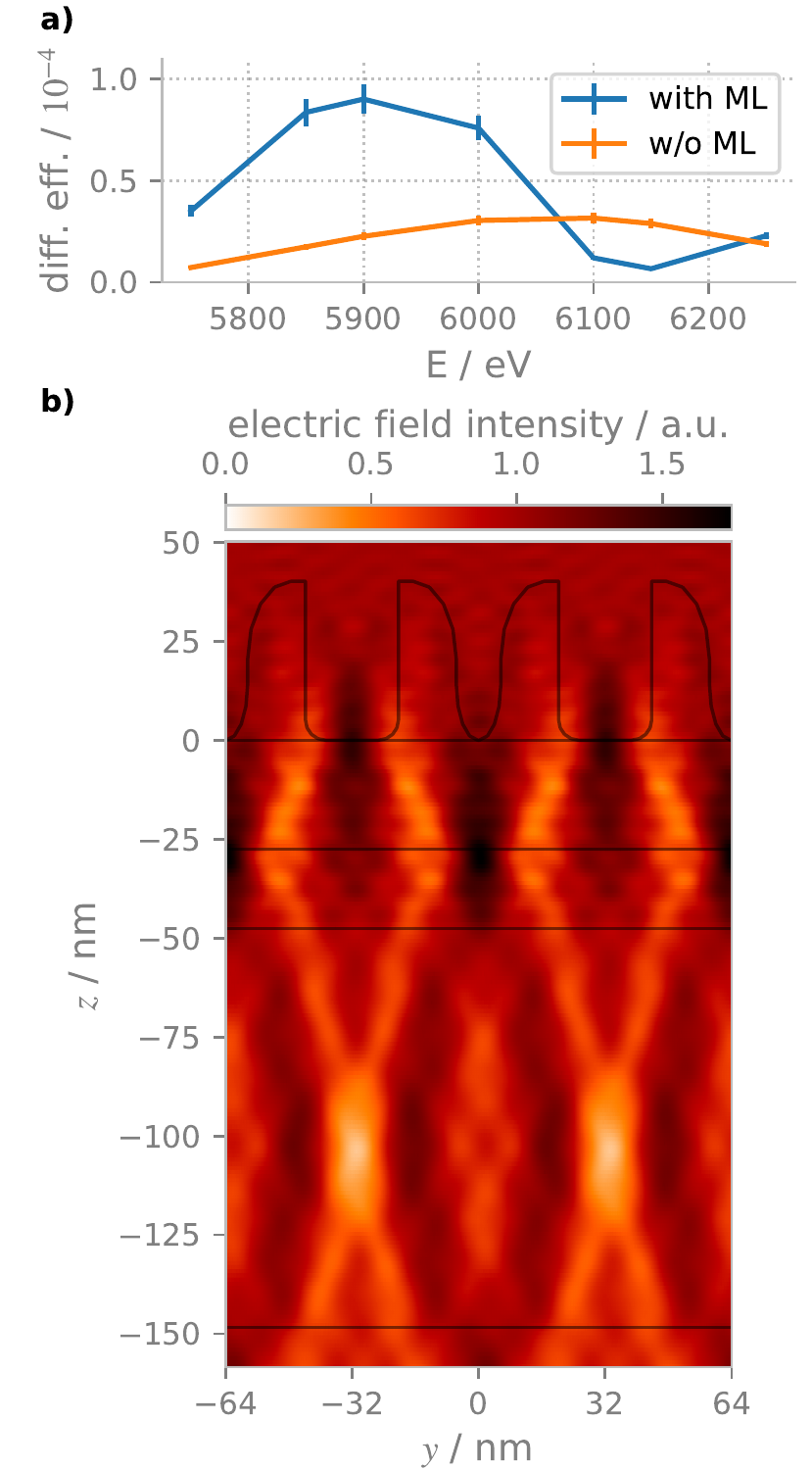}
\caption[Simulation with and without multi layer]{
a) Comparison of scattering into the first diffraction order with and without the multi layer (ML) under the grating lines.
As can be seen, the multi layer heavily influences the diffraction efficiency.
b) Simulated electric field intensity within the structure.
Black lines show the outlines of the grating lines and the boundaries of the multi layer.
The field penetrates into the layer stack and forms standing waves.}
\label{fig:noml_ml_comparison}
\end{figure}

To simulate the intensity of the diffraction orders, we solved Maxwell's equations using the finite element modelling (FEM) software package JCMsuite \cite{pomplun_2007_AdaptiveFinite} in version 3.18.15.
For this, we need to model the sample, and then discretize this model with finite elements.
For efficient FEM computations, the number of finite elements must be kept as small as possible, while still ensuring accurate results.
Fortunately, a GISAXS measurement of a grating can be reduced to a two-dimensional problem, consisting of a line profile in the $y$-$z$-plane that is extended infinitely in the $x$-direction and repeated periodically in the $y$-direction \cite{soltwisch_2017_ReconstructingDetailed}.
The line profile model we used consisted of an axially symmetric pair of lines on a substrate (see fig.~\ref{fig:line_shape_model} a).
For the reconstruction of the undisturbed line with nominally zero pitchwalk, the pitchwalk was fixed as zero, meaning that the width of the simulated unit cell was twice the pitch $2p = \SI{64}{nm}$.
Motivated by prior knowledge about the production process and electron microscopic images of cross sections of samples produced in a similar way\cite{sunday_2015_DeterminingShape}, the lines are described using the line width, the line height, elliptic rounding of three of the four corners, a side wall angle of one of the sides, and the distance between the mirrored lines.
The parameters are varied within predefined limits;
the extent of these boundaries is shown in figure \ref{fig:line_shape_model} b).
After discretizing the line profile with finite elements, Maxwell's equations are then solved for a monochromatic incident plane wave at a given incident angle.

For production reasons, there are several additional layers underneath the grating.
Unfortunately, their simulation is time-consuming due to their large height, which contributes considerably to the total size of the computational domain.
Therefore, it is tempting to neglect the effect of this multi layer by placing the grating structure directly on top of the silicon substrate in the computational model.
However, comparing the results from otherwise identical calculations with and without the multi layer (see fig.~\ref{fig:noml_ml_comparison} a), we find that the multi layer is needed for a faithful description.
The reason for this is that the incidence angle used $\alpha_i \approx \SI{1}{\degree}$ is higher than the critical angle of total external reflection $\alpha_c \approx \SI{0.3}{\degree}$;
thus, the X-rays penetrate the layer stack (see fig.~\ref{fig:noml_ml_comparison}~b).
We therefore included the multi layer in our model, with the thicknesses of the individual layers as additional parameters.

The calculation yields diffraction efficiencies $\eta$ for each diffraction order, assuming a perfect grating.
To account for the roughness of the modeled grating along the lines, a Debye-Waller like factor\cite{mikulik_1999_XrayReflection} $\exp(-q_y^2 \sigma^2)$ is introduced, with the root-mean-square roughness $\sigma$.
Additionally, to account for the X-ray intensity lost due to the aforementioned footprint effect, a loss factor $f<1$ is introduced, yielding a final intensity $I$ of:
\begin{equation}
I = \eta \, f \, e^{-q_y^2 \sigma^2} \quad.
\end{equation}
With this setup, the simulation of a single measurement geometry took about \SI{3}{s}.

With the exception of the commercial JCMsuite software package, all code used to generate the results and figures can be freely accessed and executed through Code Ocean \cite{pfluger_2020_ExtractingDimensionala}.

\section{Reconstruction of the Undisturbed Line Shape}
\begin{figure*}[t]
\includegraphics[width=\textwidth]{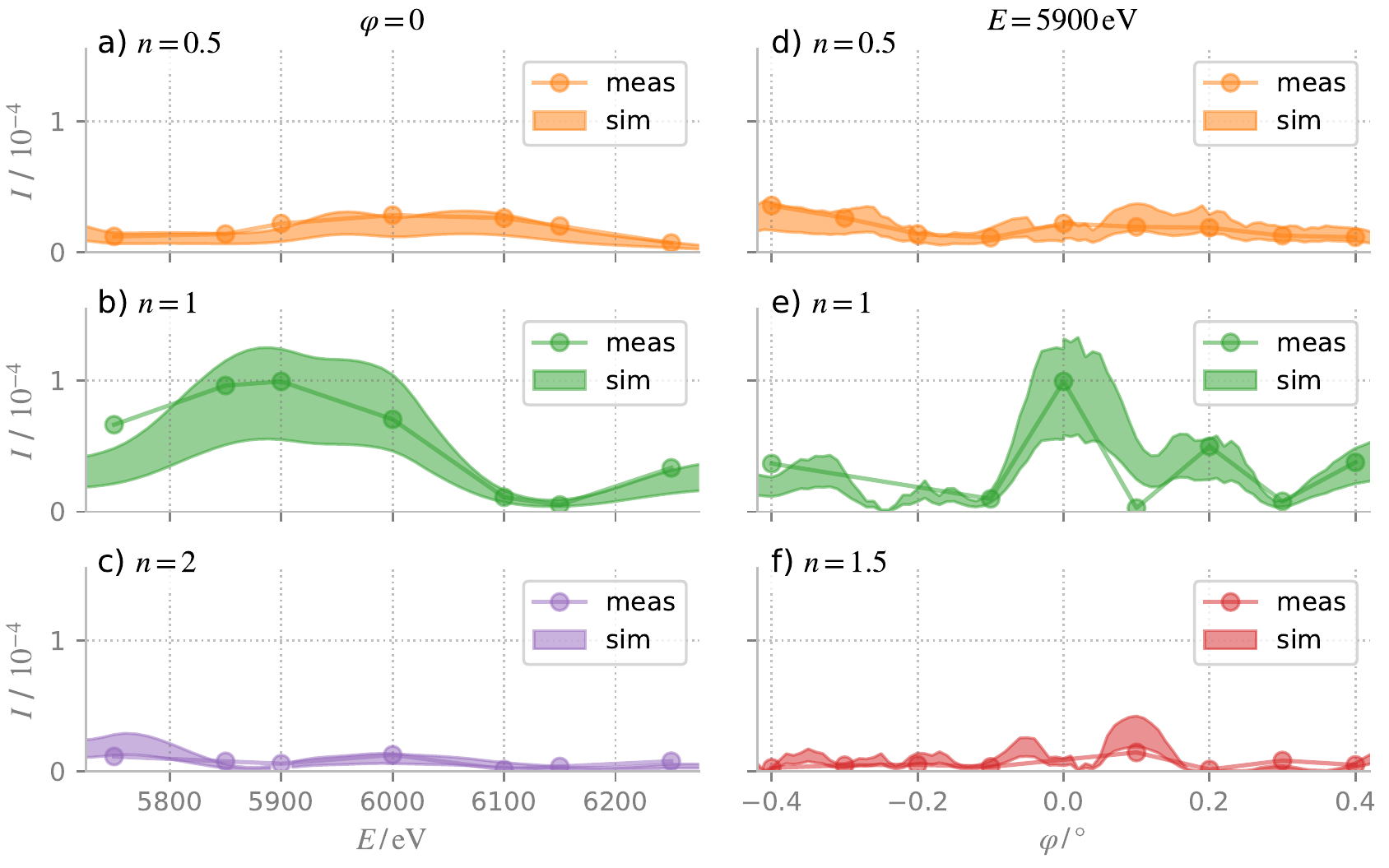}
\caption[Best fit of simulation to data]{
Best fit of simulation to data.
For clarity, only a representative subset of the data is shown.
Measured intensity $I_m$ is shown with connected circles, the corresponding simulation $I_s$ with a shaded area which represents the fitted uncertainty of the simulation.
a-c) Photon energy scan taken at $\varphi=0$, the diffraction orders with (a) $n=0.5$, (b) $n=1.0$ and (c) $n=2.0$ are shown.
d-f) $\varphi$ scan at $E=\SI{5900}{eV}$, the diffraction orders with (d) $n=0.5$, (e) $n=1.0$ and (f) $n=1.5$ are shown.
Missing data points occur when the respective diffraction order falls fully or partly into a detector gap (compare fig.~\ref{fig:gisaxs_example}~a).}
\label{fig:best_fit_r7}
\end{figure*}

We reconstructed the line shape of target PQ~4, whose nominal pitchwalk was zero, from the GISAXS diffraction intensities measured.
To recover the shape from the GISAXS measurements, the model parameters needed to be fitted to the data by minimizing the difference between the measured and simulated intensities.
For a given set of parameters, 24 measurement geometries with differing $E$ and $\varphi$ were simulated and the residual difference between the simulated and measured diffraction order intensities was calculated using the diffraction orders ranging from $n=-2$ to $n=3$ with $2n \in \ZZ$.
To minimize this residual difference, the differential evolution fitting algorithm\cite{storn_1997_DifferentialEvolution,wormington_1999_CharacterizationStructures,hannon_2016_AdvancingXray} from the SciPy software package\cite{jones_2001_SciPyOpen} was used.
The fit converged after about \num{22000} function evaluations.

To obtain uncertainties for the reconstructed parameters, the affine-invariant Markov chain Monte Carlo (MCMC) algorithm \cite{goodman_2010_EnsembleSamplers} implemented in the emcee software package \cite{foreman-mackey_2013_EmceeMCMC} was used.
For the MCMC evaluation, we use the likelihood function
\begin{align}
\mathcal{L} &= \prod_k \frac{1}{\sqrt{2 \pi u_k^2}} \exp\left(\frac{-(I_{s,k} - I_{m,k})^2}{2 u_k^2} \right) \quad,
\end{align}
with the product over all measurement points $k$, the simulated intensity of the $k$th point $I_{s,k}$, the measured intensity of the $k$th point $I_{m,k}$, and the total uncertainty of the $k$th point $u_k$.
Because not all aspects of the experimental setup can be simulated, not only the measured data, but also the simulation carries an uncertainty \cite{soltwisch_2017_ReconstructingDetailed,fernandezherrero_2019_ApplicabilityDebyeWaller}.
The uncertainty of the simulation $u_{s,k}$ is estimated using a linear error model
\begin{equation}
u_{s,k} = a \, I_{s,k} \quad ,
\end{equation}
with the error parameter $a$.
Together with the measurement uncertainty $u_{m,k}$, the total uncertainty is
\begin{equation}
u_k^2 = u_{s,k}^2 + u_{m,k}^2 \quad .
\end{equation}

When the geometrical parameters $\vec p$ of the line shape are changed, a new simulation has to be carried out to compute the likelihood.
However, if only the uncertainty factor $a$, the loss factor $f$ or the line roughness $\sigma$ are changed, the likelihood can be computed without recomputing $I_s$.
We take advantage of this fact by computing optimal values of $a$ and $\sigma$ for a given set of geometrical parameters $\vec p$ using a gradient fit, obtaining a modified likelihood function
\begin{equation}
\mathcal{L}'(\vec p, f) = \max_{a, \sigma} \mathcal{L}(\vec p, f, a, \sigma) \quad ,
\end{equation}
which we use for our MCMC evaluation.
This reduces the number of parameters in our MCMC evaluation and therefore reduces computational effort.
We nevertheless included the loss factor as a parameter in the MCMC evaluation to be able to enforce $f < 1$.

Slightly disturbed positions around the best fit from the differential evolution algorithm were utilized as the starting positions of the MCMC evaluation.
The first \num{225000} function evaluations were discarded as burn-in and the chain was run for over \num{500000} further function evaluations after the burn-in.
The best fit from the MCMC run is shown in figure \ref{fig:best_fit_r7}.
It reproduces the major features of the data measured, specifically, the relative intensities of the diffraction orders and the frequencies of the intensity oscillations in $E$ and $\varphi$.
However, the fitted uncertainty of the simulation is $a \approx \SI{39}{\percent}$, likely because our model did not include incident beam divergence due to the high computational cost of evaluating beam divergence \cite{fernandezherrero_2019_ApplicabilityDebyeWaller}.

\begin{figure*}[t]
\includegraphics[width=\textwidth]{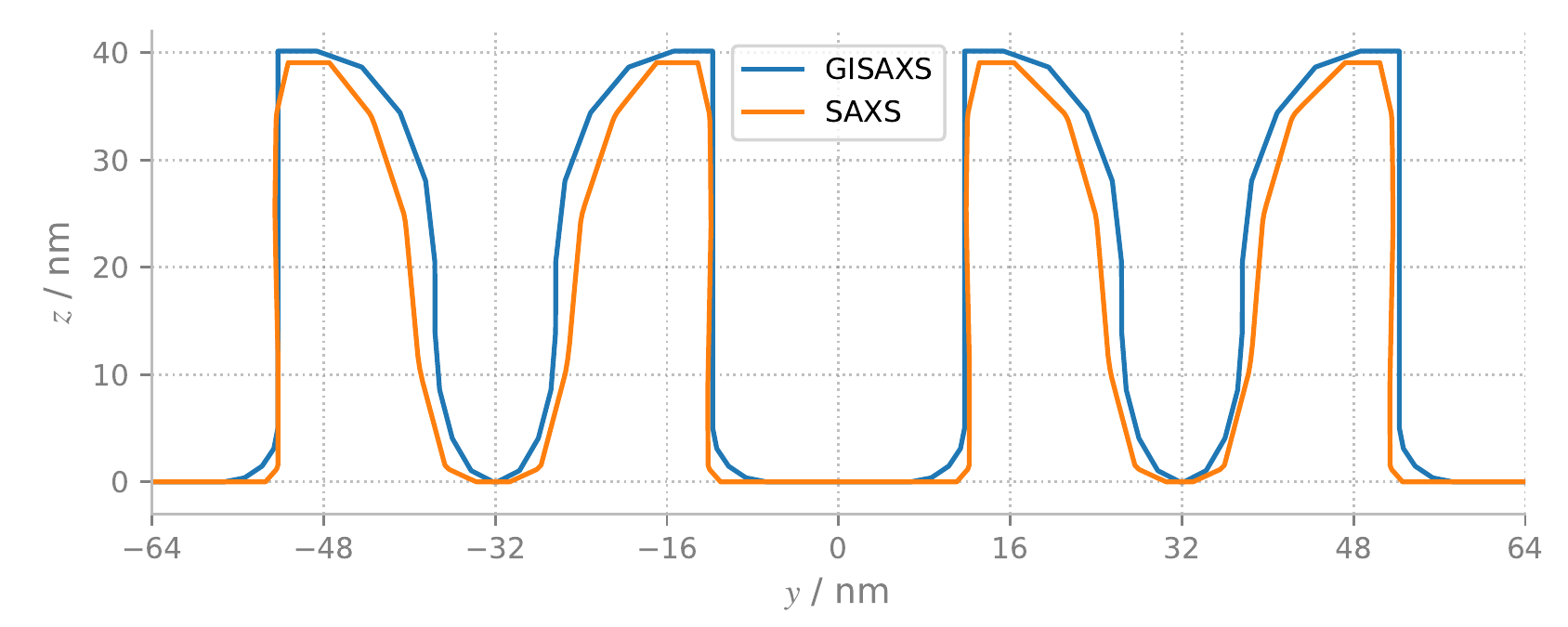}
\caption[Best fit profile]{
Best fit profile.
The fitted profile from SAXS measurements \cite{sunday_2015_DeterminingShape} of the same sample is shown for comparison.
Two unit cells of the GISAXS model are shown.
This equals one unit cell of the SAXS evaluation, which included two different line widths for the two pairs of lines.
}
\label{fig:profile_saxs_gisaxs_comparison}
\end{figure*}

The geometry of the best fit is shown in figure \ref{fig:profile_saxs_gisaxs_comparison}.
For comparison, the profile reconstructed from SAXS measurements\cite{sunday_2015_DeterminingShape} is shown as well.
Note that in the SAXS reconstruction, a model with two different line widths for adjacent pairs of lines was used.
As can be seen, the GISAXS and SAXS reconstructions agree remarkably well in terms of the general form of their lines, including the corner rounding and the slope of the walls.
However, the width and height of the lines in the reconstructions do not agree.
To quantitatively compare the GISAXS measurements with the SAXS measurements, the \SI{95}{\percent} confidence intervals were calculated from the MCMC results and compared to those extracted from the SAXS measurements.
Due to the different parametrization of the GISAXS and SAXS line shape models, only the line height and the line width (defined as the width at a height of \SI{20}{nm}) are directly comparable;
the results are shown in table \ref{tab:fitted_patameters}.
Considering the large uncertainty of the simulation, the uncertainty of the line height as reconstructed from GISAXS is remarkably small.
However, the results of SAXS and GISAXS reconstructions do not agree within their uncertainties, with a difference of \SI{1.0+-0.4}{nm} (expanded $k=2$ uncertainty).
For the line width, the uncertainty of the GISAXS reconstruction is much larger than the uncertainty of the SAXS reconstruction, and GISAXS yields a larger line width, with a difference of \SI{2.0+-1.3}{nm} (expanded $k=2$ uncertainty) compared to the average of the two line widths measured by SAXS\@.

\begin{table}[ht]
\caption[Comparison of key parameters]{\label{tab:parameters_saxs_gisaxs_comparison}
Comparison of key parameters reconstructed from GISAXS and SAXS\cite{sunday_2015_DeterminingShape} measurements.
The expanded $k=2$ uncertainties are shown.}
\label{tab:fitted_patameters}
\begin{tabular}{lcc}\toprule
Parameter & {GISAXS result} & {SAXS result} \\
          & / \si{nm}       & / \si{nm}     \\  \midrule
Line height & $40.1 \pm 0.3$ &  $39.1 \pm 0.3$ \\
Line width A & $14.4 \pm 1.3$ & $12.4 \pm 0.2$  \\
Line width B & N/A & $12.5 \pm 0.2$  \\ \bottomrule
\end{tabular}
\end{table}

\pagebreak

The total resources necessary to compute a full reconstruction are governed by the Markov chain Monte Carlo evaluation, which requires a total computation time of around one year on a single CPU core.
Utilizing the highly parallel nature of the problem and distributing the computation over several workstations, we were able to finish a full reconstruction in about one week.
Due to the significant resources necessary to compute the reconstruction, a full reconstruction of the five other measured targets with a non-zero pitchwalk was not feasible using the method presented.
In the next section, we will therefore develop an approach based on the reconstruction of the target with a nominal pitchwalk of zero to obtain measurements for the other targets much more quickly.

\section{Pitchwalk}

\begin{figure}[t]
\includegraphics[width=\columnwidth]{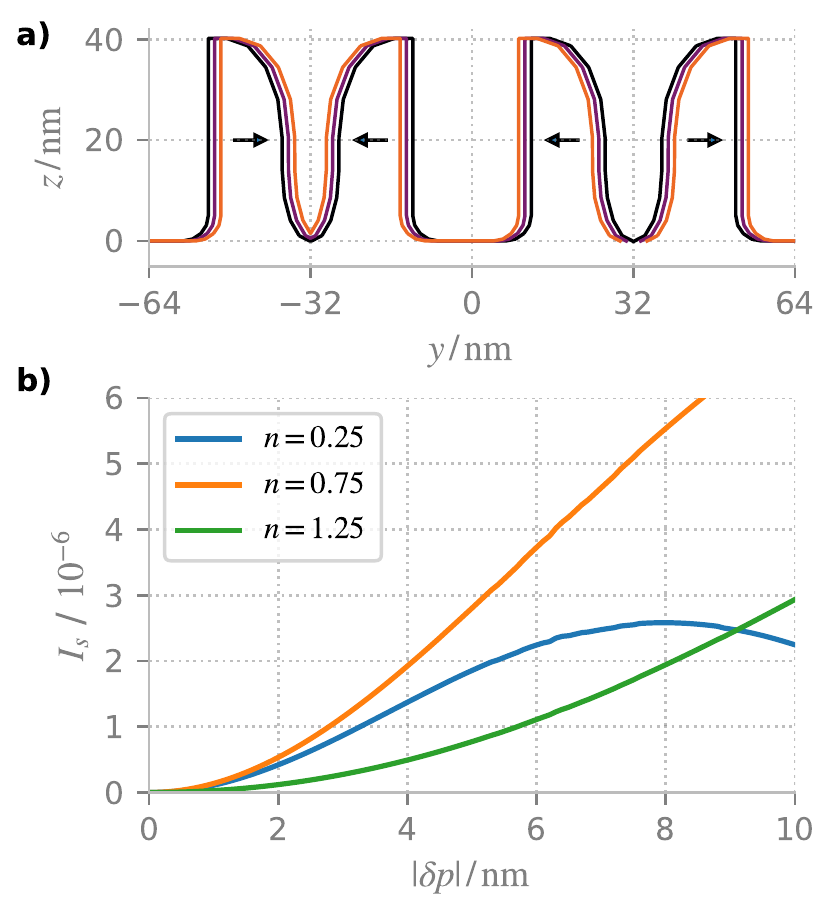}
\caption[Pitchwalk simulation library]{
Pitchwalk simulation model and results.
a) For the pitchwalk simulation, the unit cell is \SI{128}{nm} wide and one pair of lines is moved closer while the other pair is moved further apart.
b) The simulated scattering efficiencies of the quartered diffraction orders are shown.
For clarity, only the results for $E=\SI{6000}{eV}, \varphi=0$ are shown.
}
\label{fig:pitchwalk_library}
\end{figure}

To introduce the pitchwalk $\delta p$ into our computational model, it was necessary to simulate a unit cell with a width of quadruple pitch $4p = \SI{128}{nm}$.
The pitchwalk was then described by alternating the distance between pairs of lines (see fig.~\ref{fig:pitchwalk_library} a), resulting in the emergence of additional quartered diffraction orders $4n \in \ZZ$.
As a first approximation, we assumed that the shape of the lines was not affected by the pitchwalk,
and that we would therefore be able to reuse the line shape reconstructed from the undisturbed result, thus leaving the pitchwalk as the only geometrical parameter.
We calculated the diffraction order intensities for $|\delta p| \in [0, 10] \si{nm}$, in steps of \SI{0.1}{nm}, yielding a library of results (see fig.~\ref{fig:pitchwalk_library} b).
Due to the axis-symmetric nature of the problem, negative and positive pitchwalks yielded the same result, so we restricted our calculation to the magnitude of the pitchwalk $|\delta p|$.

To determine the $|\delta p|$ of a measured target, the intensity of a quartered diffraction order that arises between the main diffraction orders ($|n| = 0.25, 0.75$, or $1.25$) was compared to the simulated intensities of the diffraction order in the result library, and the $|\delta p|$ at which the difference is minimized was determined.
This yields a measurement of $|\delta p|$ for each quartered diffraction order and each measurement geometry, for a total of $N>40$ measurements per target.
We then estimated $|\delta p|$ and its type A uncertainty \cite{jcgm_2008_GuideExpression} $u(|\delta p|)$ from the arithmetic mean and the experimental standard deviation, respectively.

\begin{figure}[b!]
\includegraphics[width=\columnwidth]{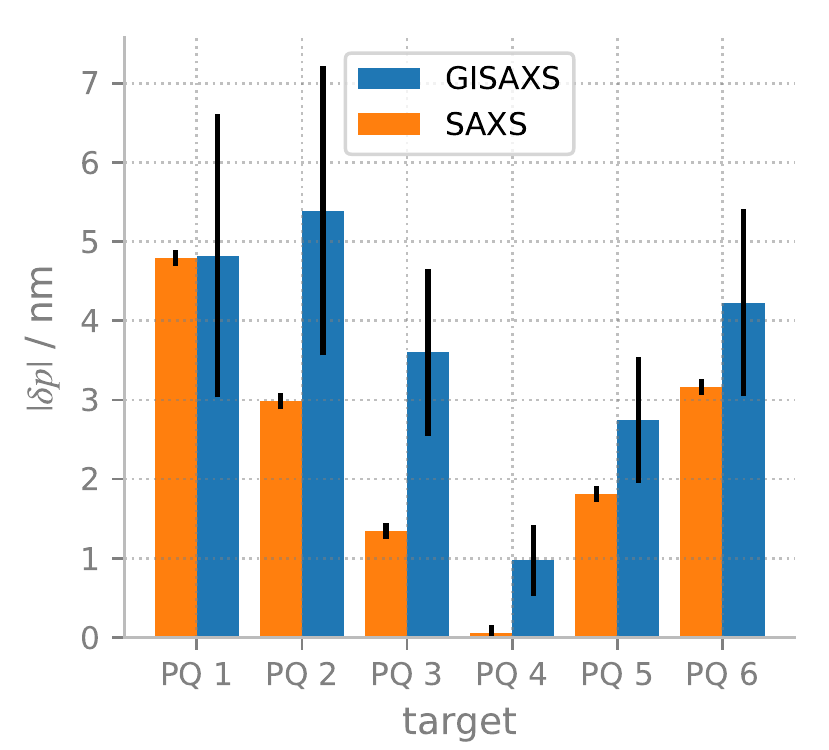}
\caption[Comparison between measurements of pitchwalk using SAXS, GISAXS]{
Comparison between measurements of pitchwalk using SAXS, GISAXS.
The black bars indicate the standard $k=1$ uncertainties.
Both measurement techniques qualitatively measure the same behavior.
The GISAXS measurements show consistently higher pitchwalk, but are compatible with the SAXS measurements due to the relatively large GISAXS uncertainties.}
\label{fig:pitchwalk_saxs_gisaxs_comparison}
\end{figure} 

The results are shown and compared to SAXS measurements in figure \ref{fig:pitchwalk_saxs_gisaxs_comparison}.
Qualitatively, the results agree, with maximum deviations between the measurements of about $2 \, u(|\delta p|)$.
This shows that, using a library approach based on a sample known to be good, pitchwalk excursions can be quantified with GISAXS measurements without the need for a time-consuming full line shape reconstruction, albeit with large uncertainties compared to SAXS\@.
However, there are also further effects that can be seen in the data.

First, the GISAXS results show a clear bias towards higher values of $|\delta p|$.
This can be explained by the secondary effects of the pitchwalk introduced that we neglected in our model due to the computational constraints.
For the SAXS measurements, a full reconstruction of the line profile was performed for all targets, which was possible because the computation of the SAXS models required much fewer recources.
In this reconstruction, it was found that the introduced pitchwalk also changed the heights and in particular the widths of the different lines in the unit cell \cite{sunday_2015_DeterminingShape}.
The change in line widths breaks the strict \SI{64}{nm} periodicity and therefore contributes to the intensity of the quartered diffraction orders in addition to the contribution of the pitchwalk.
Since we only considered the direct effect of the pitchwalk in the GISAXS model and neglected the change in line profile, our model consistently overestimates the pitchwalk to fit the observed higher intensities.

Second, at the highest $|\delta p|$ values (for sample PQ~1), the relative uncertainty of the GISAXS measurement increases considerably and the $|\delta p|$ measured is not higher than that of sample PQ~2, as would be expected.
This is likely due to the rather large changes in the line profile introduced by the highest pitchwalks.
According to the SAXS measurements, the line height of the PQ~1 sample is circa \SI{1}{nm} greater than the line height of the PQ~4 sample we used as a reference.
As the GISAXS measurements are very sensitive to changes in the line height, this deviation from our assumption of an undisturbed line shape disturbs the $|\delta p|$ determination based on the library approach, leading to diverging measurements and consequently high uncertainties.

\section{Conclusion}

Gratings manufactured using current semiconductor production techniques exhibit complex line profiles and material compositions, and perturbations such as pitchwalk might be introduced during the production process.
We have shown that both complex line profiles and pitchwalk can be reconstructed using GISAXS measurements.
However, a number of additional challenges in both the measurement and the analysis have to be overcome compared to earlier measurements \cite{soltwisch_2017_ReconstructingDetailed} of simpler samples.

The measured grating targets were surrounded by other structures, and the scattering of the surroundings contributed to the total signal.
Therefore, we suppressed the parasitic signals by using a small beam and relatively high incident angles, and included the residual parasitic signals as an additional measurement uncertainty in our further analysis.

The data acquisition and the reconstruction of the grating profiles were complicated by the small pitch $p=\SI{32}{nm}$.
The small pitch leads to relatively few grating orders ($|n| \le 2$) being scattered above the horizon even at the relatively high incident angles ($\alpha_i \approx \SI{1}{\degree}$) we used.
Nevertheless, using measurements at a range of photon energies and sample rotations allowed us to collect enough data points to successfully reconstruct the grating line profile.
The reconstructed grating profile is compatible with reconstructions from SAXS measurements within the uncertainties in the general shape, side-wall angle and corner rounding measurements, but the line widths and heights measured do not agree.
This reconstruction shows the usefulness and limitations of GISAXS as a metrology tool for small-pitch line gratings with complex line profiles.
However, the reconstruction required significant computational resources due to the larger unit cell and the multi layer under the grating, both leading to a large computational domain.
Due to limited computational resources, beam divergence could not be simulated accurately, which leads to high simulation uncertainties and consequently higher uncertainties in the geometrical parameters.

To enable fast analysis of key parameters for multiple samples despite the considerable computational resources required, we took a library approach.
By calculating a library of diffraction efficiencies from grating profiles disturbed only by pitchwalk, we could efficiently analyze a series of measurements of 6 measurement targets with varying pitchwalk.
The analysis yielded uncertainties $u(|\delta p|) < \SI{0.5}{nm}$ for the smallest pitchwalk, and higher uncertainties up to $u(|\delta p|) \approx \SI{2}{nm}$ for the highest pitchwalk.
We compared the results of our analysis with SAXS measurements, and found that the differences were $< 2.5 \, u(|\delta p|)$ for all measurement targets.
However, we also identified a bias towards systematically higher pitchwalks, and attributed this together with the higher uncertainties at high pitchwalks to additional changes in the line profile due to the pitchwalk introduced.
To improve the accuracy of the pitchwalk measurements, a more comprehensive library would be required, not only with differing pitchwalk, but also with differing line height and line widths.
Unfortunately, even including a moderate amount of additional parameters (e.g.~two line heights and two line widths) leads to an unfeasibly large number of geometries that have to be calculated for a full library (e.g.~$100^5=\num{10000000000}$).
Therefore, the development of more efficient simulation methods geared specifically to GISAXS measurements of periodic structures would be most welcome and is a field of future studies.

\section{Disclosures}
The authors declare to have no relevant financial interests in the manuscript and no other potential conflicts of interest.

\section{Acknowledgments}
We would like to acknowledge the support of Levent Cibik (PTB) in the preparation of the experiments, and Sebastian Heidenreich, Nando Farchmin, Anna Andrle, and Maren Casfor Zapata (all PTB) for fruitful discussions.

%%%%% References %%%%%

\section{References}
\printbibliography[heading=none]

%%%%% Biographies of authors %%%%%

%\listoffigures

%\listoftables

\end{document}